\documentclass[twocolumn,showpacs,preprintnumbers,amsmath,amssymb]{revtex4}

\usepackage{graphicx}
\usepackage{dcolumn}
\usepackage{bm}
\usepackage{color}
\usepackage{amssymb}

\newcommand{\dd}{{\rm d}}

\begin{document}

\preprint{ICRR-Report-585-2011-2, YITP-11-55}

\title{Open inflation in the landscape}

\author{Daisuke Yamauchi$^{1,2}$}
\email{yamauchi@icrr.u-tokyo.ac.jp}
\author{Andrei Linde$^3$}
\email{alinde@stanford.edu}
\author{Atsushi Naruko$^1$}
\email{naruko@yukawa.kyoto-u.ac.jp}
\author{Misao Sasaki$^1$}
\email{misao@yukawa.kyoto-u.ac.jp}
\author{Takahiro Tanaka$^1$}
\email{tanaka@yukawa.kyoto-u.ac.jp}
\affiliation{%
$^1$Yukawa Institute for Theoretical Physics, Kyoto University, Kyoto, Japan\\
$^2$Institute for Cosmic Ray Research, The University of Tokyo, Kashiwa 277-8582,Japan\\
$^3$Department of Physics, Stanford University, Stanford, California 94305-4060, USA
}%

\date{\today}

\begin{abstract}
The open inflation scenario is attracting
a renewed interest in the context of the string landscape.
Since there are a large number of metastable 
de Sitter vacua in the string landscape, 
tunneling transitions to lower metastable vacua 
through the bubble nucleation occur quite naturally, 
which leads to a natural realization of open
inflation.
Although the deviation of $\Omega_{0}$ from unity
is small by the observational bound, 
we argue that the effect of this small deviation 
on the large-angle CMB
anisotropies can be significant for tensor-type 
perturbation in the open inflation scenario. 
We consider the situation in which there is a large 
hierarchy between the energy scale of the quantum tunneling 
and that of the slow-roll inflation in the nucleated bubble. 
If the potential just after tunneling is 
steep enough, a rapid-roll phase appears 
before the slow-roll inflation. 
In this case  
the power spectrum is basically determined 
by the Hubble rate during the slow-roll inflation.  
On the other hand, if such a rapid-roll phase is absent, 
the power spectrum keeps the memory of the high energy density 
there in the large angular components. 
Furthermore, the amplitude of large angular components can be 
enhanced due to the effects of the wall fluctuation mode 
if the bubble wall tension is small. 
Therefore, although even the 
dominant quadrupole component is suppressed by the 
factor $(1-\Omega_0)^2$, 
one can construct some models in which the deviation of $\Omega_{0}$ 
from unity is large enough to produce measurable effects. 
We also consider a more general class of models, where the false vacuum decay may occur due to Hawking-Moss tunneling, as well as the models involving more than one scalar field. We discuss scalar perturbations in these models and point out that a large set of such models is already ruled out by observational data, unless there was a very long stage of slow-roll inflation after the tunneling. These results show that observational data allow us to test various assumptions concerning the structure of the string theory potentials and the duration of the last stage of inflation.
\end{abstract}

\pacs{Valid PACS appear here}
\maketitle

\section{Introduction}
\label{sec:introduction}

The recent observational data show that the universe is almost exactly
flat, $\Omega_{0} = 1$ with accuracy of about 1\%~\cite{Komatsu:2010fb}.
This result is in an excellent agreement with the predictions of the
simplest inflationary models \cite{Linde:2005ht}.  However, 
in the context of string landscape scenario
\cite{Kachru:2003aw,Susskind:2003kw,Freivogel:2004rd}
it is quite possible that our part
of the universe appeared as a result of quantum tunneling, after being
trapped in one of the many metastable vacua of string theory. In many
cases, tunneling can be described by a Euclidean $O(4)$-symmetric
bounce solution called Coleman-De Luccia (CDL)
instanton~\cite{Coleman:1977py,Coleman:1980aw}. Because of the symmetry of
the bounce solution, the expanding bubble has $O(3,1)$ symmetry. The
bubble formed by the CDL instanton looks from the inside like an
infinite open universe~\cite{Coleman:1977py,Coleman:1980aw}. If there
is no inflation after tunneling, the interior of the bubble becomes
an almost empty curvature dominated universe with $\Omega_{0} \ll 1$,
which is ruled out by observations. However, if the universe experienced
a sufficiently long stage of slow-roll inflation inside the light cone 
emanating from the center of the bubble, the
universe almost exactly  flat, with $\Omega_{0} \approx
1$~\cite{Gott:1982zf,Gott:1984ps,Sasaki:1993ha,Ratra:1994vw,Bucher:1994gb,Sasaki:1994yt,Lyth:1995cw,Yamamoto:1995sw,Bucher:1995ga}.

If inflation inside the bubble is too long, the universe will be
absolutely flat, and hence 
we would be unable to tell whether our part of the
universe was produced by this mechanism.  Therefore one could argue that
it would require significant fine-tuning to produce an open universe
with a minuscule but still observable deviation of $\Omega_{0}$ from
$1$, e.g. $1-\Omega_{0} \sim 10^{-2} - 10^{{-3}}$. 
However, this
conclusion is not necessarily correct. Freivogel et al. argued that
long stage of inflation in string theory may be improbable, so it
would be natural to have $\Omega_{0} \ll 1$~\cite{Freivogel:2005vv}.
The only reason why we do not live in such a universe is that 
the formation of galaxies would be
disrupted by expansion of a curvature dominated open
universe, just as it is disrupted by the
existence of a cosmological constant~\cite{Weinberg:1987dv}. This
provides a possible anthropic explanation of the smallness of the
parameter $1-\Omega_{0}$. According to \cite{Freivogel:2005vv}, the
probability of observing the deviation from flatness $1-\Omega_{0}<0.02$
is about  
 $90\%$. This result was further strengthened in \cite{DeSimone:2009dq},
 where it was shown that in a certain class of the probability measures,
 the probability of finding a universe with $1-\Omega_{0} > 10^{-3}$ is
 only about 6\%. One should bear in mind that such estimates strongly
 depend on the choice of the probability measure, which remains an
 unsolved problem. However, these estimates suggest that one can obtain
 an explanation of the observed flatness of the universe compatible with
 a relatively short stage of slow-roll inflation inside an open
 universe.
Therefore, it is not very improbable that $1-\Omega_{0}$ may be in the
observable range.  The goal of our paper is to study possible
observational consequences of the open universe scenario with
$1-\Omega_{0} \sim 10^{-2} - 10^{-3}$ for the cosmic microwave
background (CMB) anisotropy. We will focus on tensor perturbations. 

One should note that even though the basic idea of this scenario is
pretty simple, it is not easy to find a realistic open inflation model
in the single-field inflationary scenario. For example, the simplest
potential which allows inflation at $\phi >M_{\rm pl}$ and has a
metastable vacuum state at large $\phi$ can be written as a sum
$m^{2}\phi^{2}/2 - \delta\phi^{3}/3 + \lambda\phi^{4}/4$, where
$m^{2},\delta, \lambda > 0$. An investigation of inflation in this
scenario demonstrated that $V'' < H^{2}$ in the vicinity of the
barrier~\cite{Linde:1995xm}. In this case the tunneling occurs not
through the barrier, but to the top of the
barrier, described by the so-called Hawking-Moss
instanton~\cite{Hawking:1981fz,Jensen:1983ac,b60,Linde:1991sk,Batra:2006rz}.
Then we will have eternal inflation at the top of the barrier with $V' =
0$, $V'' < H^{2}$, which leads to generation of density perturbations
$O(1)$ on the curvature scale. Therefore
if the subsequent stage of the slow-roll inflation is short, we would
see enormously large CMB perturbations on the horizon. This would
dramatically contradict the results of the cosmological observations,
unless the subsequent stage of inflation is very long, which makes the
universe flat, $\Omega = 1$.

It is possible to overcome this problem and find single-field models
where CDL instantons lead to tunneling and subsequent inflation. 
Up to now, only one explicit model of this type was
proposed~\cite{Linde:1998iw,Linde:1999wv}, but the inflaton potential of
this model is quite complicated. 
The scalar perturbations ignoring the self-gravity of these fluctuations
in single-field models 
has been considered in \cite{Sasaki:1993ha,Ratra:1994vw,Bucher:1994gb,Sasaki:1994yt,Lyth:1995cw,Yamamoto:1995sw,Bucher:1995ga}.
As for the tensor perturbations, the studies in which the perturbation
of the scalar field was neglected were done in \cite{Tanaka:1997kq,Sasaki:1997ex,Bucher:1997xs,Hertog:1999kg}.
The general formula for the power spectrum for the single-field 
open inflation was given in \cite{Garriga:1998he,Garriga:1997wz}.

It is much simpler to construct models with two fields, 
where the stage of the slow-roll governed by one of the fields
is synchronized by the tunneling of another field, see
e.g.~\cite{Linde:1995xm,Yamamoto:1996qq,Sasaki:1996qn,Green:1996xe,GarciaBellido:1997te,Garriga:1997ht, GarciaBellido:1998wd,GarciaBellido:1997uh,GarciaBellido:1997hy}.
However, synchronization of the tunneling in these models is only
approximate. Therefore the geometry of the universe inside the bubble
is not exactly described by the metric of an open homogeneous
universe. That is why these models were called
 ``quasiopen''~\cite{Linde:1995xm}. 

In the simplest models of this type, the universe
inside the bubble has an islandlike structure, since a sufficiently 
large number of e folds is possible only around 
the maximum of density formed by the statistical fluctuation. 
If we consider the simplest possibility 
that there is no direct interaction between the tunneling 
field and the inflaton field that drives inflation inside the bubble, 
these models are troubled with large
supercurvature perturbations, 
especially when the energy scales before 
and after tunneling are significantly different. 
The angular power spectrum
for the supercurvature perturbations is estimated 
as~\cite{Sasaki:1996qn,Yamamoto:1996qq,GarciaBellido:1998wd}
(see also \cite{Lyth:1995cw})
\begin{eqnarray*}
\frac{\ell (\ell +1)}{2\pi}C_{\ell}^{(\Lambda )}
\approx\frac{\kappa}{\epsilon_{\rm inf}}
\left(\frac{H_{\rm L}}{2\pi}\right)^{2}
\left( 1-\Omega_{0}\right)^{\ell}
\,,
\end{eqnarray*}
where 
$\kappa =1/M_{\rm pl}^2=8\pi G$, $H_L$ is the Hubble rate on 
the false vacuum side, and 
$\epsilon_{\rm inf}$ denotes the slow-roll parameter 
during the standard slow-roll inflation.  
(In the following, we often use the Planck units in which $M_{\rm pl}$
is set to unity.) 
This amplitude of fluctuation has steep $\ell$ dependence 
and is much larger than the 
ordinary slow-roll power spectrum determined by the Hubble 
rate during the open inflation unless $1-\Omega_0$ is 
extremely small. Therefore we need to introduce some complications 
even in two field models, 
in order to make them compatible with observations~\cite{Sasaki:1996qn}. 

Originally, the models of open inflation 
were proposed in order to describe an
inflationary universe with $\Omega_{0} \sim 0.3$. 
The possibility that $\Omega_{0} \sim 0.3$ was eventually 
ruled out from the CMB observations,  
but CMB observations do not rule out the possibility that 
inflation after tunneling is sufficiently long and $\Omega_{0}$ is
sufficiently close to $1$. 
This fact gives a revived interest in the generation of perturbation in 
these models from the point of view of testing some of
the predictions of the string landscape scenario. Indeed, 
we will find that some models based on CDL tunneling 
are consistent with the existing observational data for
$1-\Omega_{0} \sim 10^{-3}$, but may be ruled out for $1-\Omega_{0} \sim
10^{-2}$. Furthermore, models of quasiopen inflation, as well as the
models leading to the Hawking-Moss
tunneling~\cite{Hawking:1981fz,b60,Linde:1991sk}, may lead to very
specific predictions, which may allow us to test basic principles of the
string inflation scenario and to study observational consequences of the
decay of the metastable string theory vacua. 
The general requirement to this class of model is that $1-\Omega_0$ must be extremely small, 
which means that inflation after the tunneling must be very long.

The models to be studied in this paper will be quite generic, but we will
keep in mind some of the expected features of tunneling in the landscape
and the last stage of the slow-roll inflation. In particular, a typical
vacuum energy in the landscape can approach Planck density or density
close to the scale of the grand unification, even though of course there
are many vacua with much smaller energy density. Meanwhile, in many
models of slow-roll inflation based on the Kachru-Kallosh-Linde-Trivedi scenario of
stabilization of stringy vacua~\cite{Kachru:2003aw}, the maximal Hubble
constant at the last stage of the slow-roll inflation cannot be greater
than the gravitino mass \cite{Kallosh:2004yh}. In such models, the typical vacuum energy of the decaying metastable vacuum in the
open universe scenario is many orders of magnitude greater than the
energy density during the last stage of the slow-roll inflation.

In this paper we will concentrate on the investigation of the models
with the CDL tunneling and investigate the CMB temperature
fluctuations. We will focus on tensor perturbations since tensor
perturbations
are more sensitive to the value of the Hubble parameter and properties
of tunneling than scalar perturbations. 
At the end of the paper we will briefly discuss scalar perturbations and their cosmological implications.

This paper is organized as follows.
In Sec. II we describe the background evolution of the models of 
open inflation in the context of the string landscape scenario. 
There we introduce models with a rapid-roll phase after the bubble
nucleation. In Sec. III we first review the method to evaluate 
the tensor perturbation after the false vacuum decay, and apply it to
our models to identify the typical signature in the CMB
spectrum. Section IV is devoted to summary.

\section{Background evolution of the open inflation models in the landscape}
\label{sec:open inflation}

\subsection{Bubble nucleation}
\label{sec:existence of rolling down phase}

\begin{figure}[tbp]
 \begin{center}
  \includegraphics[width=80mm]{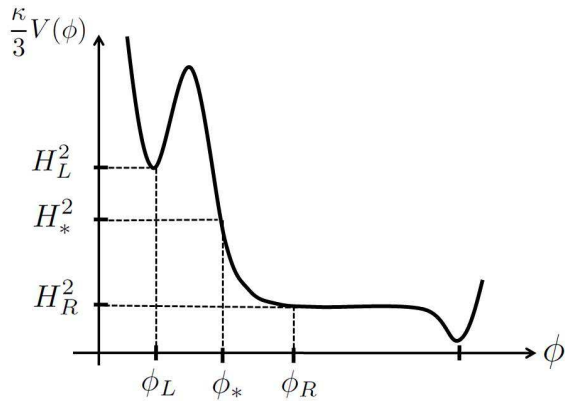}
 \end{center}
 \caption{Schematic picture of the effective potential for inflaton in one-bubble open inflation scenario.}
 \label{fig:open_potential}
\end{figure}

The system which we are going to consider consists of
a minimally coupled scalar field, $\phi$ with Einstein gravity. 
As we mentioned in Sec~\ref{sec:introduction},
the most important cosmological 
consequence in the string landscape was that there 
was a tunneling event in our past.
Let us consider an effective potential $V(\phi )$
with a local minimum at $\phi_{\rm L}$,
a slow-roll inflationary plateau at $\phi_{\rm R}$
and a point where the field emerges at $\phi_{*}$,
as shown in Fig.~\ref{fig:open_potential}.
The action is given by
\begin{eqnarray}
S=\int \sqrt{-g}\dd^{4}x
\Biggl[\frac{1}{2\kappa}R
-\frac{1}{2}g^{\mu\nu}\partial_{\mu}\phi\partial_{\nu}\phi
-V(\phi )\Biggr]\,. 
\end{eqnarray}

In this paper, we assume that the final tunneling
transition occurs through the CDL instanton.
Let us consider the $O(4)$-symmetric bounce solution~\cite{Coleman:1977py,Coleman:1980aw}. 
An $O(4)$-symmetric bubble nucleation is described by 
the Euclidean solution (instanton). 
The metric is given by 
\begin{eqnarray}
ds^2=\dd t_{\rm E}^{2}
+a_{\rm E}^2(t_{\rm E} )\left( \dd\chi_{\rm E}^{2}
+\sin^{2}\chi_{\rm E} \dd\Omega^{2}\right)\,,
\end{eqnarray}
and the background scalar field is denoted 
by $\phi =\phi (t_{\rm E})$.

The Euclidean background equations are given by
\begin{eqnarray}
&&\left(\frac{\dot{a}_{\rm E}}{a_{\rm E}}\right)^{2}
-\frac{1}{a_{\rm E}^2}
=\frac{\kappa}{3}\left( \frac{1}{2}\dot{\phi}^2
-V(\phi )\right)
\,,\label{eq:Friedmann eq}
\\
&&\left(\frac{\dot{a}_{\rm E}}{a_{\rm E}}\right)^{\cdot}
+\frac{1}{a_{\rm E}^2}
=-\frac{\kappa}{2}\dot{\phi}^2
\,,
\label{eq:Friedmann2}
\\
&&\ddot{\phi}+3\frac{\dot{a}_{\rm E}}{a_{\rm E}}\dot{\phi} 
-V'(\phi )=0
\,,\label{eq:scalar field eq}
\end{eqnarray}
where a dot represents differentiation with respect to $t_{\rm E}$. 

\begin{figure}[tbp]
 \begin{center}
  \includegraphics[width=80mm]{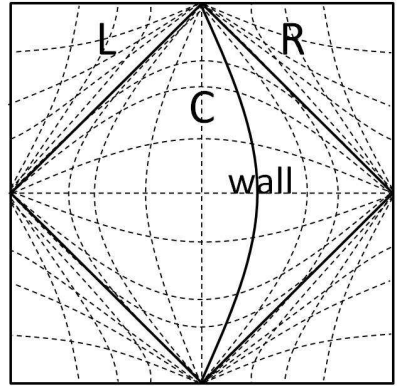}
 \end{center}
 \caption{The Penrose diagram for the bubble nucleating universe when 
the inside of the bubble terminates in a de Sitter vacuum.}
 \label{fig:open_conformal_chart}
\end{figure}

The background geometry and the field configuration
in the Lorentzian regime are obtained by the analytic
continuation of the bounce solution.
It is convenience to use the coordinate $\eta$
defined by $\dd t =a(\eta )\dd\eta$.
The coordinates in the Lorentzian regime are given by
(see e.g. \cite{Yamamoto:1996qq,Sasaki:1994yt})
\begin{eqnarray}
&&\eta_{\rm E}=\eta_{\rm C}
=-\eta_{\rm R}-\frac{\pi}{2}i
=\eta_{\rm L}+\frac{\pi}{2}i
\,,
\label{eq:eta relation}
\\
&&\chi_{\rm E}=-i\chi_{\rm C}+\frac{\pi}{2}
=-i\chi_{\rm R}
=-i\chi_{\rm L}\,,\\
&&a_{\rm E}=a_{\rm C}=ia_{\rm R}=ia_{\rm L}\,.
\end{eqnarray}
The Penrose diagram for this open Friedmann-Robertson-Walker universe is presented 
in Fig.~\ref{fig:open_conformal_chart}.

Typically, CDL instantons exist only if $V''>\kappa V$~\cite{Jensen:1983ac}. 
This condition, in combination with the requirement of the subsequent
long stage of the slow-roll inflation, requires a very special choice of
potentials in the single-field
models~\cite{Linde:1998iw,Linde:1999wv}. Alternatively, one may consider
multifield models~\cite{Linde:1995xm,Sasaki:1996qn,
Green:1996xe,GarciaBellido:1997hy,GarciaBellido:1997uh}.

In this paper, we will concentrate on the single-field models and assume
that the CDL instanton exists. 
Meanwhile, inflation typically requires $V''\ll \kappa V$.
It is difficult to make these two conditions compatible. 
Therefore one may expect that the slow-roll inflation does
 not start immediately after the CDL tunneling, 
and there must be some intermediate stage of 
rapid rolling down~\cite{Linde:1998iw,Linde:1999wv}.

This intermediate regime may last for a long time if 
the field potential is steep after tunneling. 
In general, this may be a typical situation during the tunneling
in the string landscape scenario. As we mentioned in the introduction,
the tunneling in the landscape may occur from a metastable vacuum with a
very high value of energy density. Unless we have some specific
knowledge about this vacuum state, one may assume that the scale of 
the energy density of this vacuum may be $O(1)$ in Planck units, or 
it may take some value on the grand unification energy scale 
$\sim 10^{-10}$. Meanwhile in many
models of the slow-roll inflation based on string theory with vacuum
stabilization, the Hubble constant must be smaller than the gravitino
mass~\cite{Kallosh:2004yh}. Although it is possible to avoid this
conclusion, it is very hard to do it without fine-tuning. If the
gravitino mass is of the order $1$ TeV or lower, the corresponding
vacuum energy density during the last stage of the slow-roll inflation
should be smaller than $10^{-30}$, in Planck units. 

Our knowledge of the string landscape is very incomplete, so one
should not fully rely on the estimates given above. However, on the
basis of our present understanding of the situation, one should not be
surprised to have a very long-lasting noninflationary stage after 
tunneling, which should last until the large energy density of the
metastable vacuum state is reduced by many orders of magnitude and the
slow-roll inflation begins. 
We will show that the behavior in the phase of the rapid rolling down  
significantly affects the CMB power spectrum.
In the next subsection, we will investigate 
the field evolution inside the open universe in 
the region R in Fig.~\ref{fig:open_conformal_chart}.

\subsection{Inflation in the open universe}
\label{sec:evol}

\begin{figure}[tbp]
 \begin{center}
  \includegraphics[width=80mm]{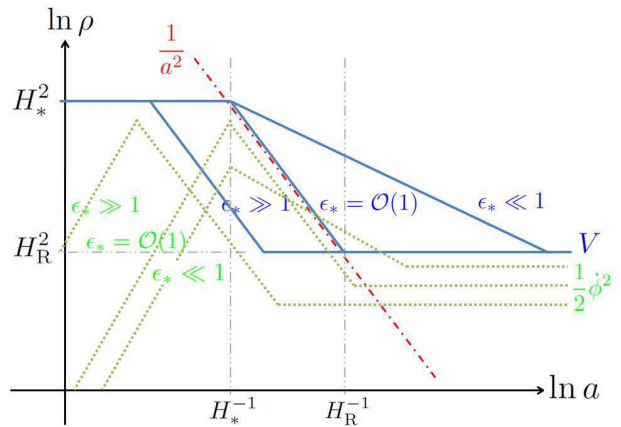}
 \end{center}
 \caption{A schematic picture of the evolution of the spatial curvature term
(the dashed-dotted line in red), the kinetic term of the field
(the dotted curves in green), and the potential term of the field
(the solid curves in blue).
}
 \label{fig:evol}
\end{figure}

In order to study the field dynamics inside the light cone emanating 
from the center of the bubble (region R), 
it is useful to use the following identity:
\begin{eqnarray}
\frac{\dd\ln\rho_{\phi}}{\dd\ln a_{\rm R}}
=-3\left( 1+w_{\phi}\right)
\,,
\end{eqnarray}
where $\rho_{\phi}=\dot{\phi}^2/2+V$, $p_{\phi}=\dot{\phi}^2/2-V$
and $w_{\phi}\equiv p_{\phi}/\rho_{\phi}$.
The asymptotic boundary conditions at the nucleation point are given by 
\begin{eqnarray}
&&a_{\rm R}(t)=t\,,\ \ \ 
\dot{\phi}(t)=-\frac{V'(\phi_{*})}{4}t\,.
\end{eqnarray}
Thus, we have
\begin{eqnarray}
1+w_{\phi}={\cal O}\biggl(\frac{\dot{\phi}^2}{V}\biggr)
={\cal O}\left(\epsilon_{*}H_*^2t^2\right)
\,.\label{eq:CD 1+w behavior}
\end{eqnarray}
where we have introduced 
\begin{equation}
\epsilon\equiv {1\over 2\kappa}\left(\frac{V'}{V}\right)^{2}\,,
\label{slow-roll0} 
\end{equation} 
and $\epsilon_{*}=\epsilon (\phi_{*})$.  
We have also introduced $H_*^2\equiv \kappa V(\phi_{*})/3$, 
which is not the expansion rate at $t=0$. 
This suppressed decreased rate of the energy density at around 
$t=0$ is caused by the large Hubble friction 
due to the spatial curvature term. 
Hence, the field at $t\approx 0$ rolls slowly even if the potential is steep.
As the spatial curvature term decays, eventually 
the potential energy starts to dominate or the energy density starts 
to decay. The potential dominance starts at $t\approx H_*^{-1}$ while
the decay of the energy density starts when $1+w_{\phi}$ becomes $O(1)$,
i.e. at $t\approx \epsilon_{*}^{-1/2}H_*^{-1}$. 
Thus, the scenario changes depending on whether 
$\epsilon_{*}$ is large or small compared with unity.
It is shown in Fig.~\ref{fig:evol} schematically 
how the evolution of the energy density depends on $\epsilon_{*}$, 
assuming that $\epsilon$ stays constant until the field 
falls down to a plateau of the potential. 

First, we consider the case with $\epsilon_{*}\ll 1$. 
In this case potential energy starts to dominate at $t\approx H_*^{-1}$ 
after a short curvature dominated stage, and there is no rapid-roll 
regime. Once the potential term dominates, 
$1+w_{\phi}$ becomes ${\cal O}(\epsilon)$ as usual.  
Then, the energy density of the field decays slowly unless the 
potential becomes steep again at a later stage. 

In contrast, if the potential is steep
at the nucleation point, i.e. $\epsilon_{*}\geq 1$, 
the situation is different.
As one can see from Eq.~\eqref{eq:CD 1+w behavior}, 
$1+w_{\phi}$ becomes ${\cal O}(1)$ at 
$t\approx \epsilon_{*}^{-1/2}H_*^{-1}$ before the potential starts to
dominate. The energy density of the field starts 
to decay rapidly proportional to $a_{\rm R}^{-3(1+w_{\phi})}$.
During this rapid-roll phase, $1+w_{\phi}$ 
is estimated as
\begin{eqnarray}
1+w_{\phi}={\cal O}\left(\epsilon\frac{\kappa V}{H^2}\right)\,.
\end{eqnarray}
where we have used $\dot{\phi}={\cal O}(V'/H)$. 
One can see that the rapid-roll phase lasts 
as far as $\epsilon\gg H^2/\kappa V(\gtrsim 1)$.
When the field falls down to a plateau of the potential, 
the velocity of the field eventually decreases. 
Then, the slow-roll inflation 
starts and the curvature term becomes completely irrelevant for 
the background expansion of the universe. 
We denote the Hubble parameter at the onset of this slow-roll 
phase by $H_{\rm R}$.

An interesting feature of the dynamics for a steep slope is ``tracking'',
which means that the ratios among the kinetic term, the potential term 
and the curvature term are approximately constant. 
Suppose $\epsilon (\phi )$ is a constant for simplicity. 
This is exactly realized when the potential is exponential-type:
\begin{eqnarray}
\frac{\kappa}{3}V(\phi)
=H_*^2\exp\Bigl[\sqrt{2\kappa\epsilon_{*}}(\phi -\phi_{*})\Bigr]\,.
\end{eqnarray}
To see that an exact tracking solution exists in this case, 
we solve Eqs.~\eqref{eq:Friedmann eq},
\eqref{eq:Friedmann2} and \eqref{eq:scalar field eq}
under the assumption 
\begin{eqnarray}
\phi (t)\propto \ln t\,. 
\end{eqnarray}
Only if $\epsilon_{*}>1$,
can one solve these equations consistently to find 
\begin{eqnarray}
&&a_{{\rm R},{\rm track}}(t)
=\frac{t}{\sqrt{1-1/\epsilon_{*}}}
\,,\label{eq:exponential sol1}\\
&&\phi_{\rm track}(t)
=\phi_{*}
+\sqrt{\frac{2}{\kappa\epsilon_{*}}}
\ln\biggl[\sqrt{\frac{3}{2}\epsilon_{*}}H_*t\biggr]\,.
\label{eq:exponential sol2}
\end{eqnarray}
It is obvious that the above solution is an exactly tracking solution.
Furthermore, one can show that it is an attractor solution.

We should note that the tracking occurs even for 
$\epsilon_{*}={\cal O}(1)$. 
The difference from the case with $\epsilon_{*}\gg 1$ is 
that the curvature term is subdominant for the expansion 
of the universe in this case during the tracking phase. 
This means that the curvature length is well outside the horizon 
scale for $\epsilon_{*}={\cal O}(1)$, while it remains 
comparable to the horizon scale for $\epsilon_{*}\gg 1$.

\begin{figure}[tbp]
 \begin{center}
  \includegraphics[width=80mm]{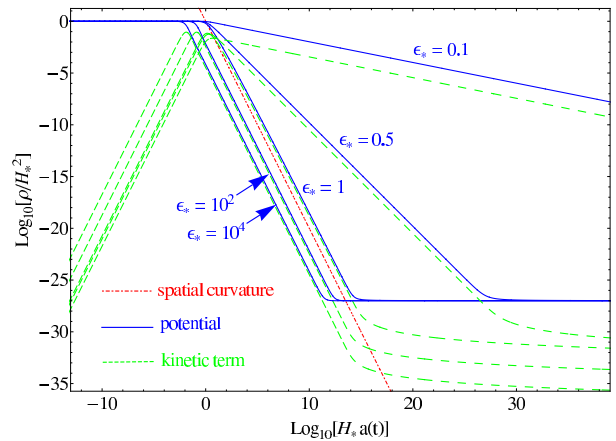}
 \end{center}
 \caption{The evolution of the energy densities in the region R 
with the exponential-type potential defined by Eq.~\eqref{eq:exponential-type}
with $\epsilon_{*}=0.1$, $0.5$, $1$, $10^2$ and $10^4$.
The dash-dotted line in red is the contribution from the spatial curvature,
the solid curves in blue are the potential energy and
the dashed curves in green are the kinetic energy.
}
 \label{fig:hybrid_evol}
\end{figure}

\begin{figure}[tbp]
 \begin{center}
  \includegraphics[width=80mm]{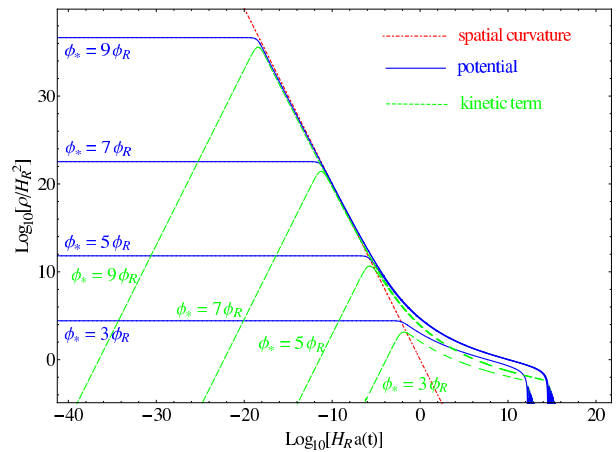}
 \end{center}
 \caption{The same as Fig.~\ref{fig:hybrid_evol},
but when the potential of the field is given by Eq.~\eqref{eq:chaotic-type}
with $\phi_{*}=9\phi_{\rm R}$, $7\phi_{\rm R}$, $5\phi_{\rm R}$, and $3\phi_{\rm R}$.}
 \label{fig:chaotic_evol}
\end{figure}

The above simplest example does not have a transition to the late-time slow-roll 
phase. Here we introduce two simple toy models, 
which have the transition from the initial rapid-roll phase 
to the late-time slow-roll phase, and we discuss the tensor perturbation 
in these models later. 

The first model has the effective potential of the following type:
\begin{eqnarray}
\frac{\kappa}{3}V(\phi )=\left( H_*^2-H_{\rm R}^2\right)\exp\Bigl[
\sqrt{2\kappa\epsilon_{*}}\left(\phi -\phi_{*}\right)
\Bigr] +H_{\rm R}^2\,,
\label{eq:exponential-type}
\end{eqnarray}
with $H_*\gg H_R$. 
The potential of a similar type often appears in
supergravity and superstring theory.
As a concrete example, we will choose the parameters 
$H_*^2=10^{-3}M_{\rm pl}^2$, $H_{\rm R}^2=10^{-30}M_{\rm pl}^2$ and
$\phi_{*}=10M_{\rm pl}$.
In Fig.~\ref{fig:hybrid_evol}, the time evolution of the energy densities for 
$\epsilon_{*}=0.1$, $0.5$, $1$, $10^2$ and $10^4$ is shown.
As is expected,
for $\epsilon_{*}<1$ the field starts to decay slowly at
$t\approx H_*^{-1}$,
for $\epsilon_{*}\geq 1$ it starts to fall down
toward the slow-roll plateau at $t=\epsilon_{*}^{-1/2}H_*^{-1}$
and its energy density decays in proportion to $1/a_{\rm R}^2$, i.e. tracking.

The second model is a variant of the potential for the 
chaotic inflation:
\begin{eqnarray}
\frac{\kappa}{3}V(\phi )=H_{\rm R}^2\left(\frac{\phi}{\phi_{\rm R}}\right)^{2}
\exp\biggl[\frac{\phi^{2}-\phi_{\rm R}^2}{\phi_{\rm R}^{2}}\biggr]\,.
\label{eq:chaotic-type}
\end{eqnarray}
The slow-roll parameter is given by 
\begin{eqnarray}
\epsilon (\phi )=\frac{2}{\kappa\phi_{\rm R}^2}
\left(\frac{\phi}{\phi_{\rm R}}+\frac{\phi_{\rm R}}{\phi}\right)^{2}
\,.
\label{slow-roll}
\end{eqnarray}
Note that we will impose $\kappa\phi_{\rm R}^2>8$ for 
the slow-roll inflation to occur at $\phi =\phi_{\rm R}$.
As with the exponential-type potential, the potential of a similar type 
also often appears in supergravity and superstring theory.
As a concrete example, we will consider the case with
$H_{\rm R}^2=10^{-40}M_{\rm pl}^2$ and $\phi_{\rm R}=7.5M_{\rm pl}$.
In Fig.~\ref{fig:chaotic_evol}, the evolution of the energy densities
for $\phi_{*}=9\phi_{\rm R}$, $7\phi_{\rm R}$, $5\phi_{\rm R}$, and $3\phi_{\rm R}$ 
is shown.
Since the field varies slowly near the nucleation point, 
the chaotic-type potential Eq.~\eqref{eq:chaotic-type}
can be approximated there by the exponential-type.
Therefore one can see that the tracking solution appears
after $t\approx \epsilon_{*}^{-1/2}H_*^{-1}$ even in this 
chaotic-type potential.

\section{CMB anisotropy in the landscape}

\subsection{Tensor power spectrum}
\label{sec: general power spectrum}

We begin with reviewing how to compute the CMB power spectrum 
due to tensor perturbation 
for single-field models of one-bubble open inflation, 
following Refs.~\cite{Garriga:1998he,Garriga:1997wz}.
Assuming that the observer is at the center of the spherical coordinates, 
the relevant tensor perturbation is the even parity one. 
We introduce the variable $U_p$ to expand the spatial metric perturbation
\begin{eqnarray}
\delta g_{ij}=a_{\rm C}^2(\eta_{\rm C})
\sum\hat{a}^{({\rm T})}_{p\ell m}
U_p(\eta_{\rm C})Y^{(+)p\ell m}_{ij}(\chi_{\rm C} ,\Omega )
+{\rm h.c.}\,,
\end{eqnarray}
where $\hat{a}^{({\rm T})}_{p\ell m}$ and $Y^{(+)p\ell m}_{ij}$ 
denote the annihilation
operator and the analytic continuation of the 
even parity tensor harmonics on the 3-hyperboloid
\cite{Tomita:1982}, respectively. 
The function $Y^{(+)p\ell m}_{ij}$ is more explicitly expressed using the
ordinary spherical harmonics $Y_{lm}$. 
In the Sachs-Wolfe formula we only need the $\chi\chi$ component, 
which is given by
\begin{equation}
 Y^{(+)p\ell m}_{\chi\chi}(\chi_{\rm C},\Omega )=
\frac{1}{\cosh^{2}\chi_{\rm C}}f^{p\ell}(\chi_{\rm C})Y_{\ell m}(\Omega)\,.
\end{equation}
Then, the time evolution equation for $f^{p\ell}(\chi_{\rm C})$ 
is in a model-independent manner given by 
\begin{eqnarray}
&&\Biggl[ -\frac{1}{\cosh^{2}\chi_{\rm C}}\frac{\dd}{\dd\chi_{\rm C}}
\cosh^{2}\chi_{\rm C}\frac{\dd}{\dd\chi_{\rm C}}
-\frac{\ell (\ell +1)}{\cosh^{2}\chi_{\rm C}}\Biggr]
f^{p\ell}
\nonumber\\
&&\qquad \qquad \qquad \qquad \qquad \ \ \ \ \ \ \ \
=\left( p^2+1\right) f^{p\ell}\,.
\end{eqnarray}

As a natural vacuum state after tunneling, we adopt 
the Euclidean vacuum, which is specified by requiring  
that the positive frequency functions are all regular at $\chi_{\rm C}=0$.
Then, we have
\begin{eqnarray}
&&f^{p\ell}(\chi_{\rm C})
=\sqrt{\frac{\Gamma (ip+\ell +1)\Gamma (-ip+\ell +1)}{\Gamma (ip)\Gamma (-ip)}}
\nonumber\\
&&\ \ \ \ \ \ \ \ \ \ \ \ \ 
\times
\frac{1}{\sqrt{\cosh\chi_{\rm C}}}P^{-\ell -1/2}_{ip-1/2}\left( i\sinh\chi_{\rm C}\right)\,,~~~~
\end{eqnarray}
where $P^{\mu}_{\nu}$ is the associated Legendre
function of the first kind.
We have fixed the normalization constant so that 
the analytic continuation of 
$Y^{p\ell m}(\chi_{\rm C},\Omega )\equiv 
f^{p\ell} (\chi_{\rm C})Y_{\ell m}(\Omega )$
to region R or L behaves as a spatial harmonic function
properly normalized on a unit 3-hyperboloid.

The equation for $U_p$ is the same as the one for
the massless scalar field~\cite{Tanaka:1997kq}:
\begin{eqnarray}
\Biggl[\frac{1}{a_{\rm C}^3}\frac{\dd}{\dd t_{\rm C}}a_{\rm C}^3
\frac{\dd}{\dd t_{\rm C}}+\frac{p^2+1}{a_{\rm C}^2}\Biggr] 
U_p=0\,.
\label{eq:U_p eq}
\end{eqnarray}
It is also convenient to introduce another variable ${\bm w}^p$,
which is related to $U_p$ by~\cite{Garriga:1998he,Garriga:1997wz}
\begin{eqnarray}
U_p=-\sqrt{\frac{\kappa}{p(1+p^2)\sinh\pi p}}\frac{1}{a_{\rm R}}
\frac{\dd}{\dd t_{\rm R}}\left( a_{\rm R}{\bm w}^p\right)\,.
\end{eqnarray}
Then, the spatial eigenfunction ${\bm w}^p$ satisfies
a simple equation 
\begin{eqnarray}
\Biggl[ -\frac{\dd^{2}}{\dd\eta_{\rm C}^2}
+U_{\rm T}(\eta_{\rm C})\Biggr]{\bm w}^p
=p^2{\bm w}^p\,,
\label{eq:w^p eq}
\end{eqnarray}
with
\begin{eqnarray}
U_{\rm T}(\eta_{\rm C})=\frac{\kappa}{2}{\phi'}^2(\eta_{\rm C})\,,
\label{eq:effective potential for mode func}
\end{eqnarray}
where a prime represents differentiation with respect to the conformal time.
Since the effective potential $U_{\rm T}$ is clearly positive definite,
there is no supercurvature mode (bound state) in the tensor-type perturbation.

From the boundary conditions for the background solution, 
we find that the potential $U_T$ vanishes at both boundaries of the
region C.
Hence, the asymptotic solutions of ${\bm w}^p$ are given by plane waves. 
We take the two orthogonal solutions ${\bm w}^p_{{\rm C}(\pm )}$
having the asymptotic behavior
\begin{eqnarray}
&&i{\bm w}^p_{{\rm C}(\pm )}\rightarrow 
\Bigg\{
\begin{array}{ll}
\rho_{\pm}^pe^{\pm ip\eta_{\rm C}}+e^{\mp ip\eta_{\rm C}}\,,
& (\eta_{\rm C}\rightarrow \pm\infty )\\
\sigma_{\pm}^pe^{\mp ip\eta_{\rm C}}\,,
& (\eta_{\rm C}\rightarrow \mp\infty )\\
\end{array}
\,,
\label{eq:scattering amplitude}
\end{eqnarray}
as independent solutions. 
The reflection and 
transmission coefficients satisfy the
Wronskian relations:
\begin{eqnarray}
|\rho_{\pm}^p|^2+|\sigma_{\pm}^p|^2=1\,,\ 
\sigma_{+}^p=\sigma_{-}^p\,,\ 
\sigma_{+}^p\overline{\rho_{-}^p}
+\overline{\sigma_{-}^p}\rho_{+}^p=0\,.
\label{eq:Wronskian}
\end{eqnarray}
Using the relation between the 
coordinates~\eqref{eq:eta relation},
we analytically continue Eq.~(\ref{eq:w^p eq})
to the region R.
Then, we obtain the evolution equation for ${\bm w}^p$, 
\begin{eqnarray}
\Biggl[-\frac{\dd^{2}}{\dd\eta_{\rm R}^2}
+U_T(\eta_{\rm R})-p^2\Biggr]{\bm w}^p=0\,.
\label{eq:w^p eq in R}
\end{eqnarray}
Similarly, the analytic continuations of 
the two independent modes to the region R are given by 
\begin{eqnarray}
&&{\bm w}^p_{{\rm R}(+)}
=e^{\pi p/2}\rho_{+}^p\overline{\tilde{\bm w}^p}
+e^{-\pi p/2}\tilde{\bm w}^p
\,,\\
&&{\bm w}^p_{{\rm R}(-)}
=e^{\pi p/2}\sigma_{-}^p\overline{\tilde{\bm w}^p}
\,.
\end{eqnarray}
where $\tilde{\bm w}^p$ is the 
solution of Eq.~\eqref{eq:w^p eq in R}
that asymptotically behaves like 
$\tilde{\bm w}^p\rightarrow e^{ip\eta_{\rm R}}$
in the limit $\eta_{\rm R}\rightarrow -\infty$.
The amplitude of the fluctuation is proportional to the sum 
of squares of these two independent modes, which is evaluated as 
\begin{eqnarray}
&&\bigl|{\bm w}^p_{{\rm R}(+)}\bigl|^{2}
+\bigl|{\bm w}^p_{{\rm R}(-)}\bigl|^{2}
\nonumber\\
&&\ \ 
=2\biggl[\bigl|\tilde{\bm w}^p\bigl|^{2}\cosh\pi p
+{\rm Re}\left(\rho_{+}^p\Bigl(\overline{\tilde{\bm w}^p}\Bigr)^{2}\right)
\biggr]
\,.
\end{eqnarray}

After the horizon crossing, $a^2H^2\gg p^2+1$, Eq.~\eqref{eq:U_p eq}
implies that the amplitude of $U_p$ freezes.
In terms of ${\bm w}^p$, two independent solutions 
after the horizon crossing are given by 
\begin{eqnarray}
{\bm w}^p_1
\propto a_{\rm R}^{-1}\,,\ \ \ 
{\bm w}^p_2\propto 
a_{\rm R}^{-1}\int^{\eta}a_{\rm R}^2(\eta' )\dd\eta'
\,.
\end{eqnarray}
Neglecting the decaying mode ${\bm w}^p_1$, we denote 
the asymptotic behavior of $\tilde {\bm w}^p$ as
\begin{eqnarray}
\tilde{\bm w}^p\rightarrow -{\mathbb H}_p
e^{ip\eta_{\rm T}^p}
a_{\rm R}^{-1}\int^{\eta}a_{\rm R}^2(\eta' )\dd\eta'
\,,
\label{eq:asymptotic behavior of w^p}
\end{eqnarray}
where ${\mathbb H}_p$ and $\eta_{\rm T}^p$ are the 
amplitude and phase, respectively.
We note that $p\eta_{\rm T}^p$ vanishes in the limit $p\rightarrow 0$
because $\overline{\tilde{\bm w}^p}=\tilde{\bm w}^{-p}$. 

Using ${\mathbb H}_p$ and $\eta_{\rm T}^p$, 
the even parity tensor power spectrum can be 
expressed as \cite{Garriga:1998he,Garriga:1997wz}
\begin{eqnarray}
&&\frac{p^3}{2\pi^{2}}P_{\rm T}(p)
\equiv
\frac{p^3}{2\pi^{2}}
\left(
\bigl| U_{p(+)}\bigl|^{2}
+\bigl| U_{p(-)}\bigl|^{2}
\right)
\nonumber\\
&&\ \ \ \ \ \ \ \ \ \ \ \ \ \ 
=4\kappa\left(\frac{{\mathbb H}_p}{2\pi}\right)^{2}\frac{p^2\coth\pi p}{1+p^2}
\Bigl( 1-y_{\rm T}^p\Bigr)\,,
\label{eq:general power spectrum}
\end{eqnarray}
where 
\begin{eqnarray}
&&y_{\rm T}^p=-\frac{1}{\cosh\pi p}{\rm Re}
\left(\rho_{+}^pe^{-2ip\eta_{\rm T}^p}\right)
\,,
\end{eqnarray}
essentially represents
the effects of the bubble wall. 

To summarize, 
in order to find the power spectrum for tensor-type perturbations,
one has to solve the scattering problem \eqref{eq:w^p eq}
in the region C to obtain $\rho_{+}^p$ and 
the evolution equation~\eqref{eq:asymptotic behavior of w^p}
in the region R to obtain ${\mathbb H}_p$ and $\eta_{\rm T}^p$.

Our main focus in this paper is on the latter process. 
Investigating the various possibilities for the tunneling potential 
is beyond the scope of this paper. 
We simply use the known results under thin-wall approximation~\cite{Sasaki:1997ex}. 
For a thin-wall bubble, the effective potential $U_{\rm T}$
is characterized 
by two constant parameters, $\eta_{\rm W}$ and $\Delta s$:
\begin{eqnarray}
U_{\rm T}(\eta_{\rm C})
\approx \Delta s\delta (\eta_{\rm C}-\eta_{\rm W})\,,
\label{eq:thin-wall potential}
\end{eqnarray}
It can be shown that 
$\eta_{\rm W}$ and $\Delta s$ are expressed in terms of 
the vacuum energies and 
the surface tension of the wall, 
$S_1=\int\dd t_{\rm C}\dot{\phi}^2$,
as
\begin{eqnarray}
&&\Delta s=\frac{2}{\sqrt{(\alpha +1)^2+4\beta}}
\,,\label{eq:Delta s eq}\\
&&\eta_{\rm W}=\frac{1}{2}\ln
\left(\frac{[\sqrt{(\alpha +1)^2+4\beta}+\alpha +1]^{2}}
{4\beta}\right)
\,,\label{eq:eta_W eq}
\end{eqnarray} 
with
\begin{eqnarray}
\alpha =\frac{4(H_{\rm L}^2-H_*^2)}{\kappa^{2} S_1^2}
\,,\  
\beta =\left(\frac{2H_*}{\kappa S_1}\right)^{2}\,,
\label{eq:alpha beta}
\end{eqnarray}
where $H_{\rm L}=\sqrt{\kappa V(\phi_{\rm L})/3}$ and 
$H_*=\sqrt{\kappa V(\phi_{*})/3}$.

In terms of $\eta_{\rm W}$ and $\Delta s$, 
we can solve the equation for the mode function \eqref{eq:w^p eq}
to obtain the reflection coefficient $\rho_{+}$:
\begin{eqnarray}
\rho_{+}^p=\frac{-ie^{2ip\eta_{\rm W}}\Delta s}{2p+i\Delta s}
\,.
\label{eq:reflection}
\end{eqnarray}
From Eq.~\eqref{eq:Wronskian},
the transmission probability from the false vacuum is given by
$\bigl|\sigma_{+}^p\bigl|^{2}=1/[1+(\Delta s/2p)^2]$.
From this expression, we find that 
efficiently reflected at the wall are only low frequency modes 
with $p\lesssim \Delta s/2$, 
which are thought to be interpreted as the wall fluctuation~\cite{Sasaki:1997ex}. 
Notice that the reflection rate at $p=0$ is unity. 
Without this reflection by the wall, 
the square amplitude of fluctuation is divergent.   
In this sense, one might be able to interpret that 
the so-called wall fluctuation mode represents the 
penetration of the waves from the false vacuum side, 
and its amplitude is reduced if 
the wall efficiently reflects back the waves. 
Substituting the expression (\ref{eq:reflection}), the 
spectrum becomes 
\begin{eqnarray}
&&\frac{p^3}{2\pi^{2}}P_{\rm T}(p)
=4\kappa\left(\frac{{\mathbb H}_p}{2\pi}\right)^{2}
\frac{p^2\coth\pi p}{1+p^2}
\nonumber\\
&&\ \ \times
\Biggl[ 1-\frac{(\Delta s)^2\cos [b_pp] 
+2p\Delta s\sin [b_pp]}
{[4p^2+(\Delta s)^2]\cosh\pi p}\Biggr]\,,
\label{eq:power spectrum}
\end{eqnarray}
where $b_p:=2\left(\eta_{\rm W}-\eta_{\rm T}^p\right)$.
Once parameters $\eta_{\rm W}\,,\Delta s$ and the late-time 
asymptotic value
of $\tilde{\bm w}^p$ are fixed, we can immediately
calculate the  power spectrum using this formula. 
The power spectra for the tensor-type perturbation 
are shown in Fig.~\ref{fig:constHubble}
for $(\alpha ,\beta )=(100,1)$, $(1,100)$, $(10^{-3},10^{-3})$ and
$(1,1)$, setting ${\mathbb H}_p=H_R$ 
and $\eta_T^p=0$. 
As a reference, we plot 
the spectrum for ${\mathbb H}_p=H_R$, $\Delta s=\infty$ and $b_p=0$, 
which we denote by ``plain spectrum'' in this paper,  
as a dashed curve in black in the same figure. 



\subsection{Effects of inflation in the open universe}

Now we move on to the problem of solving
the evolution equation~\eqref{eq:asymptotic behavior of w^p}
in the region R to evaluate ${\mathbb H}_p$ and $\eta_{\rm T}^p$.

\subsubsection{Slow-roll case}

\begin{figure}[tbp]
 \begin{center}
  \includegraphics[width=80mm]{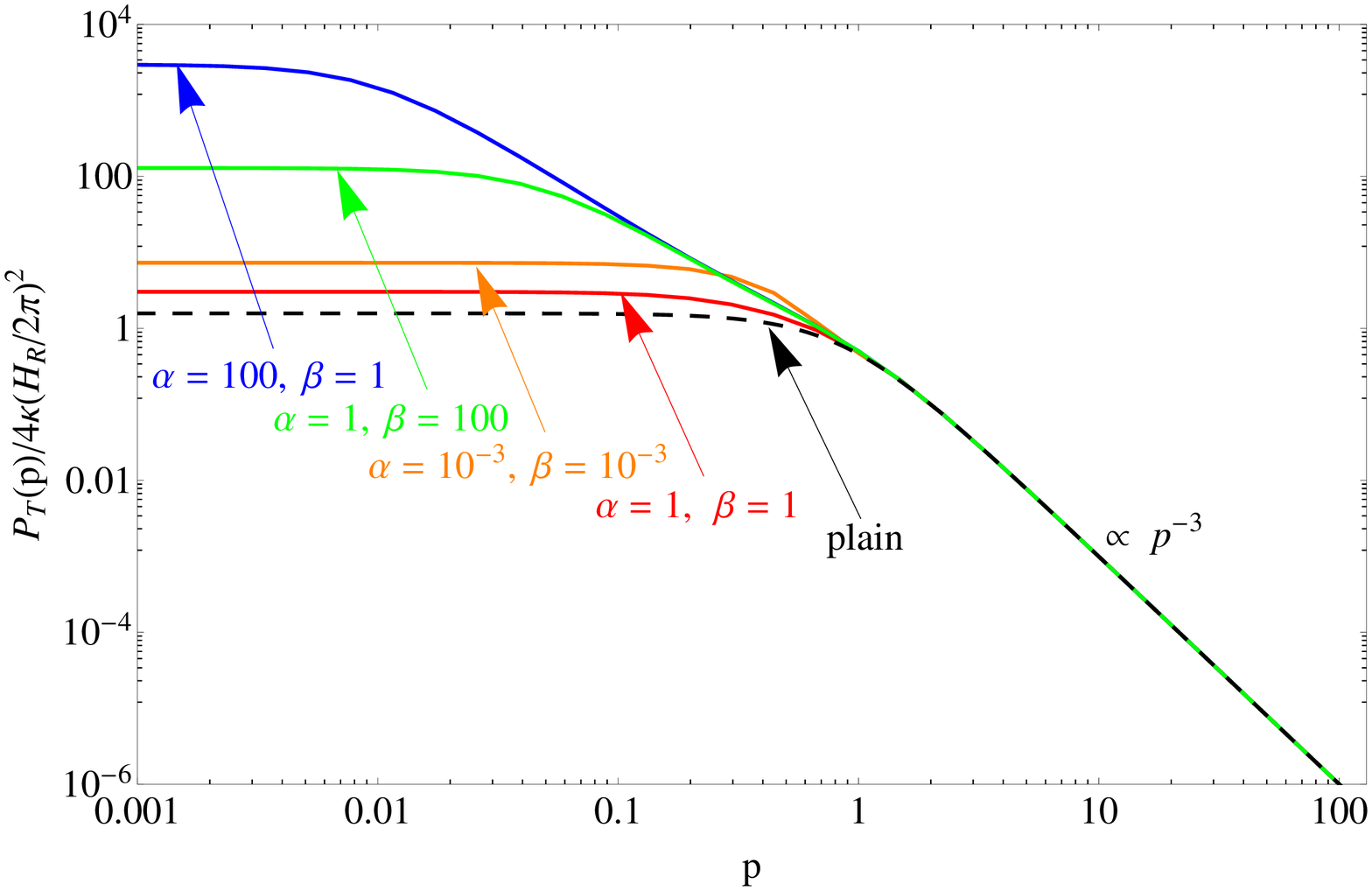}
 \end{center}
 \caption{
The power spectrum due to tensor-type perturbation.
The solid curves are, from top to bottom,  
for $(\alpha ,\beta )=(100,1)$, $(1,100)$, $(10^{-3},10^{-3})$ and
 $(1,1)$, setting ${\mathbb H}_p=H_R$ and $\eta_T^p=0$. 
The dashed curve in black is the spectrum for 
$\Delta s=\infty$ and $b_p=0$, which we call ``plain 
spectrum.''}
 \label{fig:constHubble}
\end{figure}

We begin with the case with $\epsilon_{*}\ll 1$. 
From the discussion in Sec.~\ref{sec:evol}, there is no
rapid-roll phase for $\epsilon_{*}\ll 1$, and 
the field kinetic term is always 
irrelevant for the cosmic expansion rate. 
In this case one can neglect the potential $U_{\rm T}$ in 
the field equation for the mode function ${\bm w}^p$~\eqref{eq:w^p eq in
R}. 
Hence, ${\bm w}^p$ behaves like a plane wave and 
its amplitude stays constant. 
The scale factor will also be approximated by $a_R\approx -1/H_R\eta$ in 
this case. Then, from Eq.~(\ref{eq:asymptotic behavior of w^p}), we 
find ${\mathbb H}_p\approx H_R$ and $\eta_{\rm T}^p\approx 0$.  
Namely, the plain spectrum is obtained. As seen from the formula 
(\ref{eq:power spectrum}), the amplitude of tensor perturbation is basically 
scale invariant ($P_T\propto p^{-3}$) whose amplitude is 
determined by the Hubble rate inside the bubble, except for 
the modes near $p=0$. 
For the modes with $p\alt \Delta s$, there is suppression due to 
the reflection by the wall as mentioned earlier, and hence the 
spectrum $P_T$ does not grow indefinitely for small $p$.  
In Fig.~\ref{fig:constHubble}, 
we plot the power spectrum 
due to tensor-type perturbation with 
${\mathbb H}_p=H_{\rm R}$, $\eta_{\rm T}^p=0$
and various tunneling parameters $\alpha$, $\beta$.

This result tells us that there is no memory from the high 
energy false vacuum in large $p$ modes. 
In other words, 
the high frequency modes simply stay in the adiabatic vacuum state 
irrespective of the presence of the bubble wall and the initial high 
energy vacuum. 

This can be also understood as follows. 
It will be natural to assume that 
the partial wave $U_p$ typically has the amplitude 
of $O\left(H_*/M_{\rm pl}\right)$ 
as a memory of the high energy false vacuum 
when the physical wavelength is as short as $H_*^{-1}$. 
At this initial time the scale factor will be given 
by $a_{{\rm R},{\rm init}}=\sqrt{p^2+1}/H_*$. 
On the other hand, the freeze-out of the amplitude of $U_p$ 
occurs at around the horizon crossing time, 
at which the scale factor is estimated as 
$a_{{\rm R},{\rm final}}=\sqrt{p^2+1}/H_{\rm R}$. 
As the scale factor evolves from $a_{{\rm R},{\rm init}}$ 
to $a_{{\rm R},{\rm final}}$, 
the amplitude of the tensor-type perturbation $U_p$ before the 
horizon crossing decays in proportion to $1/a_{\rm R}$. 
Hence, the amplitude frozen after the horizon crossing
would be estimated as 
\begin{eqnarray}
p^{3/2}\bigl| U_p\bigl|\,
\approx\frac{H_*}{M_{\rm pl}}\frac{a_{{\rm R},{\rm init}}}{a_{{\rm R},{\rm final}}} 
\approx\frac{H_{\rm R}}{M_{\rm pl}}\,. 
\end{eqnarray}
The amplitude of the spectrum 
is solely determined by $H_{\rm R}$ independently of 
the initial value of the Hubble parameter $H_*$. 
This rough estimate explains the model-independent behavior of the 
tensor perturbation spectrum in the high frequency region.

\subsubsection{Rapid-roll case}

\begin{figure}[tbp]
 \begin{center}
  \includegraphics[width=80mm]{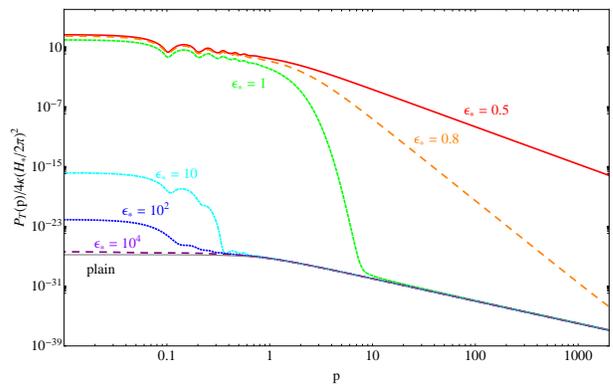}
 \end{center}
 \caption{The tensor-type power spectrum 
with the exponential-type potential Eq.~\eqref{eq:exponential-type}.
The curves are, from top to bottom, the spectrum with 
$\epsilon_{*}=0.5$, $0.8$, $1$, $10$, $10^2$ and $10^4$.
Here we set $\alpha=1$ and $\beta=1$ as typical parameters for which 
the contribution from the wall fluctuation mode is negligibly small.
For comparison, we also plot the plain spectrum with the gray solid line.}
 \label{fig:hybrid_spectrum}
\end{figure}

\begin{figure}[tbp]
 \begin{center}
  \includegraphics[width=80mm]{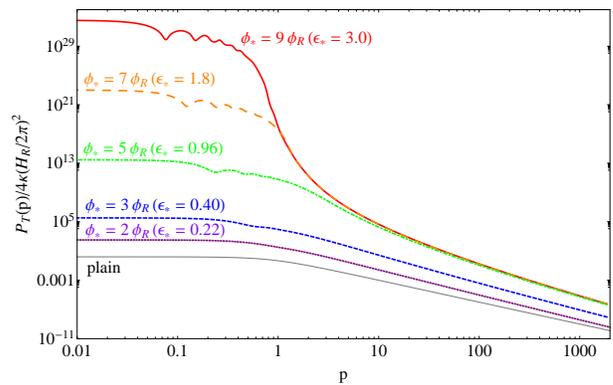}
 \end{center}
 \caption{The same as Fig.~\ref{fig:hybrid_spectrum},
but the potential of the field is given by Eq.~\eqref{eq:chaotic-type}
with $\phi_{*}=9\phi_{\rm R}$, $7\phi_{\rm R}$, $5\phi_{\rm R}$, $3\phi_{\rm R}$, and $2\phi_{\rm R}$.}
 \label{fig:chaotic_spectrum}
\end{figure}

Next we consider the models with the long-lasting rapid-roll phase 
in the region R. 
In Fig.~\ref{fig:hybrid_spectrum}, we present 
the computed tensor perturbation spectra 
for the exponential-type model defined by Eq.~\eqref{eq:exponential-type}, 
setting $\alpha=1$ and $\beta=1$ as typical parameters for which 
the contribution from the wall fluctuation mode is negligibly small.
As seen from the figure, the spectra are quite different from
the plain spectrum, for which we neglect 
 not only the contribution of the wall fluctuation mode 
but also the effect of the field dynamics in the region R. 
The power spectrum for $\epsilon_{*}={\cal O}(1)$
increases sharply for small $p$, i.e. highly red-tilted.
As we discussed in Sec.~\ref{sec:evol}, 
in this case the curvature length is well 
outside the horizon size during the tracking phase. 
Since the amplitude for small $p$ depends on 
the value of the Hubble rate at the time when 
the curvature scale starts to deviate from the 
horizon size, it keeps the memory of the high expansion rate of 
${\cal O}(H_*)$ 
at the onset of the rapid-roll phase.
Therefore the squared amplitude for small $p$ is proportional to ${\cal O}(\kappa H_*^2)$.  
In contrast, the wavelength of the modes with sufficiently large $p$ stays 
inside the horizon size during the tracking phase. 
Only after the field arrives at the slow-roll plateau do 
these modes cross the horizon.
Thus, the squared amplitude of the modes with large $p$ 
becomes ${\cal O}(\kappa H_{\rm R}^2)$.
For this reason, the spectrum 
for $\epsilon_{*}={\cal O}(1)$ is steeply red-tilted.

By contrast, if the potential is extremely steep, $\epsilon_{*}\gg
1$, at the nucleation point, 
the field starts to roll down rapidly before the curvature term 
becomes irrelevant for the expansion of the universe. 
As long as the potential continues to be steep enough,  
the curvature length stays comparable to the horizon size.  
If this remains the case until the field reaches 
the slow-roll plateau, even the amplitude of the modes with $p\alt 1$ 
is governed by the Hubble rate at the slow-roll plateau, $H_{\rm R}$.
Therefore there is no significant enhancement of the amplitude 
for small $p$ in this case. 

Finally, the spectrum for $\epsilon_{*}\ll 1$ can be understood in 
the usual manner. The spectrum is similar to 
the one presented in Fig.~\ref{fig:constHubble}, 
in which the field dynamics in the region R is neglected. 
The same plot for the chaotic-type model defined by 
Eq.~\eqref{eq:chaotic-type}
is shown in Fig.~\ref{fig:chaotic_spectrum}.
As with the exponential-type model, 
the spectra for $\epsilon_{*}={\cal O}(1)$
have large enhancement for small $p$, while the spectra
with $\epsilon_{*}<1$ are similar to 
the one presented in Fig.~\ref{fig:constHubble}. 
As the initial position $\phi_{*}$ is increased, 
the magnitude for large $p$ decreases because it is 
normalized by the Hubble
parameter at the nucleation point $H_*$, which becomes larger.

\subsection{CMB temperature anisotropy}
\label{sec:previous result}

We translate the spectrum  for tensor-type perturbation 
obtained in the preceding subsection into CMB temperature 
anisotropies, following the discussion given in Ref.~\cite{Sasaki:1997ex}.
The large-angle CMB temperature anisotropies 
due to tensor-type perturbation are simply evaluated by the 
Sachs-Wolfe formula 
\begin{eqnarray}
&&\frac{\Delta T}{T}(\hat{\bm n})
=-\frac{1}{2}\int^{\eta_{0}}_{\eta_{{\rm LSS}}}
\dd\eta \delta g^{\prime}_{ij}
\left(\eta ,x^i(\eta )\right)\hat{n}^{i}\hat{n}^{j}
\,,
\end{eqnarray}
where $\eta_{0}$ and $\eta_{\rm LSS}$, respectively, denote the conformal time 
at the present epoch and that at the last scattering surface, 
$\hat{n}^{i}$ is the unit vector along the observer's line of sight and 
$x^i(\eta )=(\eta_{0}-\eta )\ \hat{n}^i$ represents the photon trajectory.

Since we are interested in the anisotropies on large angular scales,
we consider only those modes that enter the Hubble horizon
after the universe becomes matter dominated.
Thus, we set the scale factor to $a(\eta )=\cosh\eta -1$.
The conformal time then can be written in terms 
of the density parameter $\Omega_0$ and the redshift $z$:
\begin{eqnarray}
&&\eta (z)=2{\rm arccosh}
\sqrt{1+\frac{\Omega_{0}^{-1}-1}{1+z}}
\,,\label{eq:conformal time}
\end{eqnarray}
where we have neglected the effect of the dark energy.

The evolution equation for $U$ is given by \cite{Kodama:1984}
\begin{eqnarray}
U_p''+2\frac{a^{\prime}}{a}U_p'+(p^2+1)U_p=0\,. 
\end{eqnarray}
The above equation can be solved exactly to give
\begin{eqnarray}
U_p(\eta )=U_p(0)G_p(\eta )\,,
\end{eqnarray}
where $U_p(0)$ is the initial amplitude of fluctuations 
given by Eq.~\eqref{eq:general power spectrum} and 
\begin{eqnarray}
G_p(\eta )=\frac{3\left(\cosh\left(\frac{\eta}{2}\right)
\frac{\sin p\eta}{2p}
-\sinh\left(\frac{\eta}{2}\right)\cos p\eta\right)}
{(1+4p^2)\sinh^{3}\left(\frac{\eta}{2}\right)}\,, 
\label{eq:G_p def}
\end{eqnarray}
is the growing mode function.
Now, we decompose the temperature anisotropies in terms
of the spherical harmonics:
\begin{eqnarray}
\frac{\Delta T}{T}(\hat{\bm n})
=\sum_{\ell ,m}\int^{\infty}_0\dd p
\left(\frac{\Delta T}{T}\right)_{p,\ell}
Y_{\ell m}(\hat{\bm n})\,,
\end{eqnarray}
where 
\begin{eqnarray}
\left(\frac{\Delta T}{T}\right)_{p,\ell}
\!\! Y_{\ell m}
&\!\!=\!\!&-\frac{1}{2}U_p(0)
\nonumber\\
&&\!\!
\times\int^{\eta_{0}}_{\eta_{\rm LSS}}
d\eta\, G^{\prime}_p(\eta )
Y^{p\ell m}_{\chi\chi}(\eta_{0}-\eta ,\Omega )
\,,~~
\end{eqnarray}
and the radial-radial component of 
the tensor-type harmonic function 
$Y^{p\ell m}_{\chi\chi}(\chi ,\Omega )$ is given
by 
\begin{eqnarray}
&&Y^{p\ell m}_{\chi\chi}(\chi ,\Omega )
=\sqrt{\frac{1}{2(p^2+1)}\frac{(\ell +2)!}{(\ell -2)!}}
\nonumber\\
&&\ \ \ \ \ 
\times\biggl|\frac{\Gamma (\ell +1+ip)}{\Gamma (1+ip)}\biggl|
\frac{P^{-\ell -1/2}_{ip-1/2}(\cosh\chi )}
{\sinh^{5/2}\chi}Y_{\ell m}(\Omega )\,.~~
\end{eqnarray}

The temperature anisotropies are usually expressed 
in terms of $C_{\ell}$, the multipole moments
of the temperature autocorrelation function, defined by 
\begin{eqnarray}
\Biggl\langle\frac{\Delta T}{T}(\hat{\bm n})
\frac{\Delta T}{T}(\hat{\bm n}')\Biggr\rangle
=\frac{1}{4\pi}\sum_{\ell}(2\ell +1)C_{\ell} 
P_{\ell}(\cos\theta )\,,
\end{eqnarray} 
where $\cos\theta =\hat{\bm n}\cdot\hat{\bm n}'$.
Then, the multipole moments for the 
tensor-type perturbation $C^{({\rm T})}_{\ell}$
are given by
\begin{eqnarray}
C^{({\rm T})}_{\ell}
\!\!\!&=\!\!&\int^{\infty}_0\!\! \dd p\,\Biggl\langle 
\left|\frac{\Delta T}{T}\right|^{2}_{p,\ell}\Biggr\rangle
\nonumber\\
&=\!\!&\frac{1}{8}
\frac{(\ell +2)!}{(\ell -2)!}
\int^{\infty}_0\dd p\frac{P_T(p)}{p^2+1}
\Biggl|\frac{\Gamma (\ell +1+ip)}{\Gamma (1+ip)}\Biggr|^{2}
\nonumber\\
&&\  
\times
\Biggl|\int^{\eta_{0}}_{\eta_{{\rm LSS}}}d\eta G_p^{\prime}(\eta )
\frac{P^{-\ell -1/2}_{ip-1/2}(\cosh (\eta_{0}-\eta ))}
{\sinh^{5/2}(\eta_{0}-\eta )}\Biggr|^{2}
.~~~~~
\label{eq:Clformula}
\end{eqnarray}

\begin{figure}[tbp]
 \begin{center}
  \includegraphics[width=80mm]{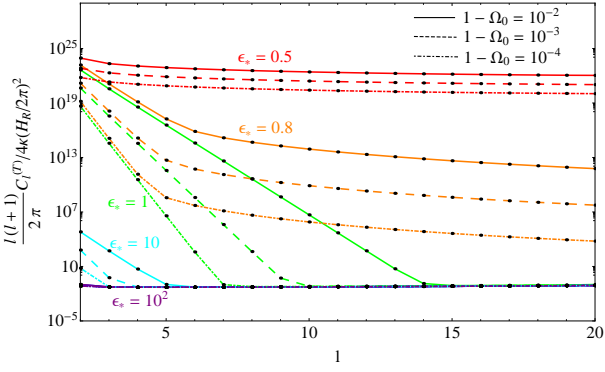}
 \end{center}
 \caption{The multipole moments for the tensor-type perturbations
for the exponential-type potential Eq.~\eqref{eq:exponential-type}
with $\epsilon_{*}=0.5$, $0.8$, $1$, $10$ and $10^2$.
Again we set $\alpha=1$ and $\beta=1$ as a representative case. 
}
 \label{fig:hybrid_ClT}
\end{figure}

\begin{figure}[tbp]
 \begin{center}
  \includegraphics[width=80mm]{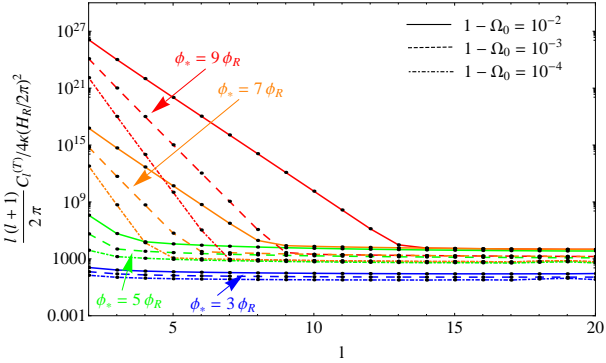}
 \end{center}
 \caption{The same as Fig.~\ref{fig:hybrid_ClT},
but when the field potential is given by Eq.~\eqref{eq:chaotic-type}
with $\phi_{*}=9\phi_{\rm R}$, $7\phi_{\rm R}$, $5\phi_{\rm R}$, and $3\phi_{\rm R}$.}
 \label{fig:chaotic_ClT}
\end{figure}

We will discuss possible observational signatures in the CMB 
temperature anisotropies in the open inflation scenario revived in 
the context of string theory landscape that prefers a 
moderately small value of $1-\Omega_{0}$. 
The value of $\Omega_{0}$ depends rather sensitively on the
 underlying scenario.
The prediction for the value of $\Omega_{0}$, based on the 
primordial distribution of the number of the e folds 
of the inflation inside the bubble and the anthropic bias, was
investigated in \cite{Freivogel:2005vv,DeSimone:2009dq}.
There is a big uncertainty because the resulting distribution of 
the value of $\Omega_{0}$ strongly
depends on the choice of the probability measure, 
but their estimates imply that it can be natural that
$1-\Omega_{0}$ is in the marginally observable range.
For this reason, we focus on the case that $1-\Omega_{0}$
is in the observable range, e.g. from $10^{-3}$ to $10^{-2}$, and 
evaluate the CMB anisotropies for the previously 
 introduced two toy models. 

The computed CMB temperature power spectra 
for the exponential-type and chaotic-type 
models are shown 
in Figs.~\ref{fig:hybrid_ClT} and \ref{fig:chaotic_ClT}, respectively.
Again we set $\alpha=1$ and $\beta=1$ as a representative case 
in which the contribution from the wall fluctuation mode is negligibly small.
One can see that the tensor CMB angular power spectrum
for small $\ell$ behaves like $(1-\Omega_{0})^{\ell}$, while it agrees 
with the scale invariant inflationary tensor spectrum for large $\ell$.
Compared with the amplitude of the tensor perturbation 
for the slow-roll inflation at $H\approx H_R$, 
there is significant enhancement for small $\ell$ 
due to the field dynamics in the region R 
for $\epsilon_{*}\alt 1$.
Hence, we find that, unless $\epsilon$ is not extremely large, 
rapid rolling down affects the CMB spectrum at low $\ell$ 
significantly. 

It is known that the spectrum Eq.~\eqref{eq:Clformula}
can be approximately decomposed into two pieces:
\begin{eqnarray}
C_{\ell}^{({\rm T})}=P_{\rm W}\,\tilde{C}_{\ell}^{({\rm W})}
+C_{\ell}^{({\rm T},{\rm res})}
\,,
\end{eqnarray}
where $P_{\rm W}\,\tilde{C}_{\ell}^{({\rm W})}$ 
represent contributions due to wall fluctuations:
\begin{eqnarray}
&&P_{\rm W}=\int^{\infty}_0\dd pP_{\rm T}(p)
\,,\label{eq:P_W def}\\
&&\tilde{C}_{\ell}^{({\rm W})}=
\frac{(\ell +2)!(\ell !)^2}{8(\ell -2)!}
\label{PW}
\nonumber\\
&&\ \ \ 
\times
\Biggl|\int^{\eta_{0}}_{\eta_{{\rm LSS}}}d\eta G_0^{\prime}(\eta )
\frac{P^{-\ell -1/2}_{-1/2}(\cosh (\eta_{0}-\eta ))}
{\sinh^{5/2}(\eta_{0}-\eta )}\Biggr|^{2}\,.
\label{eq:ClW_tilde}
\end{eqnarray}
We have also introduced $C_{\ell}^{({\rm T},{\rm res})}$, 
which corresponds to
the continuous spectrum due to standard tensor perturbations, as 
the residual piece of the spectrum. 
The previous plots presented 
in Figs.~\ref{fig:hybrid_ClT} and \ref{fig:chaotic_ClT} are 
basically corresponding to $C_{\ell}^{({\rm T},{\rm res})}$. 

The effect of the wall fluctuation mode is rather easy to investigate 
analytically, and in many cases it is more important than the 
continuum modes. 
To see observational signatures 
when the deviation of $\Omega_{0}$ from unity is small, 
it will be useful to expand quantities with respect to $1-\Omega_{0}$.
The spectrum due to the wall fluctuations given by Eq.~\eqref{eq:ClW_tilde}
reduces to 
\begin{eqnarray}
\frac{\ell (\ell +1)}{2\pi}\tilde{C}_{\ell}^{({\rm W})}
\approx\frac{(\ell +2)!(\ell +1)!}{800\pi (\ell -1)\Gamma (\ell +3/2)}
(1-\Omega_{0})^{\ell}\,.
\end{eqnarray}
Unless $1-\Omega_{0}$ is close to unity, 
the effects of wall fluctuation mode mainly appear 
in the quadrupole of CMB temperature fluctuation. 

The amplitude of the wall fluctuation mode can be also 
evaluated analytically. 
As seen in Figs.~\ref{fig:constHubble}, \ref{fig:hybrid_spectrum}, 
and \ref{fig:chaotic_spectrum},
the integral (\ref{PW}) is dominated by the contribution from $p$ around
0. Thus, we expand $P_{\rm T}(p)$ around $p= 0$ as
\begin{eqnarray}
&&\frac{1}{2\pi^{2}}P_{\rm T}(p)
\approx\frac{4\kappa}{\pi}\left(\frac{{\mathbb H}_0}{2\pi}\right)^{2}
\frac{1}{p^2+(\Delta s/2)^2}
\nonumber\\
&&\ \times
\Biggl[
1-\left(\frac{b_0\Delta s}{2}\right)\frac{\sin [b_0p]}{b_0p}
+\frac{1}{2}\left(\frac{b_0\Delta s}{2}\right)^{2}
\frac{\sin^{2}\bigl[\frac{b_0p}{2}\bigr]}{\bigl[\frac{b_0p}{2}\bigr]^{2}}
\Biggr]\,,\nonumber\\
\label{eq:P_T expand}
\end{eqnarray}
where we assumed that $p\ll 1$ but the combination 
$b_0 p$ is not necessarily small. 
Notice that $b_p=2(\eta_{\rm W}-\eta^{p}_{\rm T})$ also depends
on the field dynamics in the region R through $\eta^p_{\rm T}$, 
which can be large. 
Substituting Eq.~\eqref{eq:P_T expand} into Eq.~\eqref{eq:P_W def},
we find that the amplitude due to wall fluctuations can be
approximated by 
\begin{eqnarray}
&&\frac{P_{\rm W}}{2\kappa H_*^2}\approx
\frac{1}{\Delta s}\left(\frac{{\mathbb H}_0}{H_*}\right)^{2}
f\left({\Delta s b_0\over 2}\right)\,,
\label{eq:amp wall}
\end{eqnarray}
with 
\begin{eqnarray}
f(x)=2e^{-(x+|x|)/2}-1+x\,.
\end{eqnarray}
This expression (\ref{eq:amp wall}) is valid even if $\Delta s |b_0|$ is large.
The expression contains exponential but, in fact,  
$f(x)$ is just $x-1$ for $x\gg 1$, while 
$f(x)\approx x+1$ for $x\alt 1$. 
Therefore, $P_W$ only weekly depends on
$\eta_{\rm T}^0$. 
Roughly speaking, the above expression indicates that 
the amplitude of the wall fluctuations can be
significantly enhanced only when $({\mathbb H}_0/H_*)^2/\Delta s$ is large. 
Small $\Delta s$ can be realized for a small wall tension $S_1$ 
since $\Delta s$ is 
roughly estimated as $\Delta s\approx \kappa S_1/2H_*$ when 
$H_L$ and $H_*$ are the same order.  
An interesting observation is that
the amplitude of wall fluctuation also depends on ${\mathbb H}_0$. 
Therefore, when the continuum contribution $C_{\ell}^{({\rm T},{\rm res})}$ 
for small $\ell$ is enhanced 
due to the evolution inside the region R, the wall fluctuation 
mode is also enhanced. 

\begin{figure}[tbp]
 \begin{center}
  \includegraphics[width=80mm]{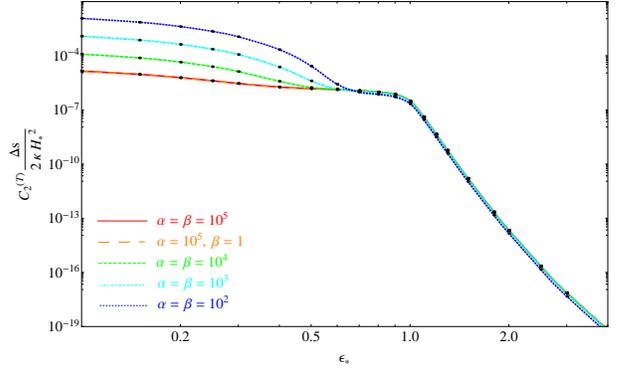}
 \end{center}
 \caption{CMB quadrupole amplitude 
as a function of $\epsilon_*$ for 
various values of $(\alpha, \beta)$:  
$(10^2,10^2)$, $(10^3,10^3)$, $(10^4,10^4)$, $(10^5,10^5)$ and $(10^5,1)$.  
They correspond to $\Delta s\approx  2\times 10^{-2}$,
$2\times 10^{-3}$, $2\times 10^{-4}$, $2\times 10^{-5}$ 
and $2\times 10^{-5}$, respectively. 
The value of $1-\Omega_0$ is fixed to $10^{-3}$. 
Here the amplitude multiplied by $\Delta s$ is shown. 
Therefore these plots clearly show that 
the quadrupole amplitude scales like $1/\Delta s$ for
 $\epsilon_*\agt 1$. 
}
 \label{fig:newfig}
\end{figure}

It is shown in Fig.~\ref{fig:newfig} how 
the amplitude of $C_2^{(T)}$ depends on $\epsilon$ for the case of 
the exponential-type model defined by Eq.~\eqref{eq:exponential-type}. 
The value of $1-\Omega_0$ is fixed to $10^{-3}$ here. 
In these plots the magnitude is rescaled by 
multiplying $\Delta s$.  This figure shows that the power for the
quadrupole increases for small $\Delta s$ precisely in proportion to $1/\Delta s$ 
for $\epsilon_*\agt 1$ as expected. 
For $\epsilon_*\alt 1$, the contribution from the continuum becomes
significantly large. As a result, the magnitude becomes larger 
for a smaller value of $\Delta s$.  

It may be noted that both modes at $p\approx 0$ and at $p\agt 1$ 
contribute to the CMB quadrupole, but the former modes
physically represent the wall fluctuation degree of freedom.
Therefore they might have a property similar to the scalar-type
perturbation. Thus these two different contributions 
might be distinguished by looking at the CMB polarization.

If a model predicts large enhancement of low $\ell$ modes 
due to tensor perturbation, such a model may contradict 
the current observation. This gives a constraint on 
the model building based on string theory. 
We are interested in the case when $H_*$
is close to Planck scale but inflation in the region R 
occurs at sufficiently low energies. 
We also require $1-\Omega_0$ is not extremely small. 
As advocated by Freivogel et al.~\cite{Freivogel:2005vv},
this may be achieved naturally in the string landscape.

Then, first of all, a very small $\epsilon_*$ is not compatible with
the above scenario because the slow-roll inflation phase 
starts before the energy density decays sufficiently.
This means that the amplitude of the tensor perturbation is 
too large.  
When $\epsilon_*\gg 1$, the contribution to the CMB spectrum 
from the tensor perturbation is largely suppressed 
by the factor $({\mathbb H}_0/H_*)^2$. 
Even if $\epsilon_*=O(1)$, there is a suppression factor 
$(1-\Omega_0)^\ell$.  On the other hand, there is a possible huge 
enhancement factor $\sim 1/\Delta s$ 
due to the wall fluctuation mode, which can be 
very large if the wall tension is small. 
However, one cannot choose a very small $\Delta s$ in 
the present context when $H_*$ is close to the Planck scale. 
An extremely small $\Delta s$ is in conflict with 
the requirement that the bubble nucleation rate is sufficiently
suppressed in order to avoid too large a signature of bubble 
collisions in our observable universe. 

The above three factors, the values of
$\epsilon_*$, $\Delta s$ and $1-\Omega_0$ 
basically control the amplitude 
of the CMB spectrum for small $\ell$ due to the
tensor perturbation, which is observationally constrained 
to be not too large. We should stress that 
the constraint from the tensor contribution can be avoided
if one can arrange the potential slope after the bubble nucleation 
to be sufficiently steep, i.e., if $\epsilon_*\gg1$. 

In this paper, we did not study how the factor coming 
from the tunneling process, which is $1/\Delta s$ in the 
thin-wall case, 
is related to the tunneling potential 
beyond the thin-wall approximation. 
For example, we did not discuss the case that 
the energy density at the nucleation point 
is already largely reduced, in spite of the high false 
vacuum energy. Such a tunneling process cannot 
be described by the thin-wall approximation. 
We will come back to this issue in our future publication.

\section{Summary}
\label{sec:summary}

The inflation scenario with CDL tunneling, 
the so-called open inflation has attracted renewed attention
in the context of string landscape.
The most important cosmological consequence in the string landscape
is that it is likely that there were a sequence of tunneling events 
in our past and our universe is produced inside a bubble.

In this paper we focused on those models
that experience the CDL tunneling followed by the last
inflation before the beginning of our hot Friedmann universe
inside a nucleated bubble. 
Our interest was in the landscape scenario in which 
the energy density at the nucleation point is 
close to the Planck scale, while a
much lower energy scale is preferred for this last inflation 
inside the nucleated bubble. 
We also assumed a moderately small $1-\Omega_0$ 
preferred from a perspective of the anthropic probability
distribution.  Under these assumptions we examined the
CMB temperature anisotropies due to the tensor-type perturbation.

We considered single-field models, 
assuming that the condition for the presence of a
CDL instanton $V''>\kappa V$ is satisfied around 
the potential barrier. 
In this case it seems difficult to imagine that  
such a potential is connected to the one that satisfies 
one of the slow-roll conditions $V''<\kappa V$ immediately 
after tunneling. Therefore it seems fairly natural to have 
some intermediate stage of rapid rolling down after tunneling.
We found that a rapid-roll phase significantly modifies
the power spectrum of the tensor-type perturbations.
In particular, the behavior of the power spectrum 
strongly depends on the slow-roll parameter at the nucleation 
point, $\epsilon_{*}$, where $\epsilon\equiv (V'/V)^2/(2\kappa)$
as defined in Eq.~(\ref{slow-roll0}).

When $\epsilon_{*}\alt 1$, there is no rapid-roll phase. 
Then, the high rate of cosmic expansion near the nucleation 
point is imprinted in the low $\ell$ components of CMB
and the spectrum becomes red-tilted. 
Such a feature would contradict the observed CMB spectrum 
if the potential energy at the nucleation point were
sufficiently high and $1-\Omega_0$ were not extremely small. 

As we noted in the introduction, if the transition were
driven by the Hawking-Moss instanton, which is the case in the inflationary models with the simplest polynomial potential $m^{2}\phi^{2}/2 - \delta\phi^{3}/3 + \lambda\phi^{4}/4$, the large-angle CMB fluctuations would be
significantly enhanced due to the scalar-type perturbations. Such models would contradict current observations, unless the last stage of inflation after the tunneling is very long. 
A similar conclusion is valid for the quasiopen scenario, which involves more than one scalar field \cite{Linde:1995xm}. This scenario seems much easier, if not easiest, to implement in string theory, as compared to other options.

Thus one of the important conclusions is that
we are already testing string theory landscape against observations, 
and we already know quite a lot about the structure of the potential, 
assuming that inflation after tunneling is rather short.
Note that this assumption ~\cite{Freivogel:2005vv} is based on a specific 
choice of the probability measure in eternal inflation, and on some additional assumptions about the inflationary potentials in the landscape. Our analysis shows that either we can rule out a large class of inflationary potentials 
which at the first glance seem quite reasonable or we can rule out the assumption
that a long stage of the slow-roll inflation is improbable.

By contrast, when $\epsilon_{*}\gg 1$, we have 
a rapid-roll phase after the bubble nucleation. 
In this case, the CMB spectrum
behaves as if there is no memory of the high energy regime. 
Namely, the magnitude of CMB spectrum is basically determined by 
the slow-roll inflation at low energies, which 
succeeds the rapid-roll phase. Therefore 
the tensor perturbations tend to be negligible, 
which is consistent with observations. We also note
that the scalar-type perturbations are expected to be more
sensitive to the evolution of the field. Since the
power spectrum for gauge-invariant spatial curvature perturbation 
will be roughly given by $(H_{\rm R}^2/\dot{\phi})^2$, 
one can expect that the amplitude of 
the scalar perturbations will be suppressed during the rapid-roll phase. 
Since this effect occurs at the first stages of inflation after the tunneling, 
it may lead to suppression of the large scale adiabatic perturbations~\cite{Linde:1998iw,Linde:1999wv}
This tendency seems to fit well with the observed suppression of the 
CMB quadrupole moment~\cite{Contaldi:2003zv}.

The above conclusion is slightly modified by 
the presence of the wall fluctuation modes. 
When the wall tension $S_1$ is much smaller than 
$M_{\rm pl}\sqrt{\rho_*}$, where $\rho_*$ is the energy 
density at around the tunneling, the wall fluctuation modes 
enhance the low $\ell$ components in the CMB spectrum 
in proportion to $S_1^{-1}(1-\Omega_{0})^{\ell}$.
Therefore a possible observational signature of 
the wall fluctuation modes is basically in the CMB quadrupole.
The amplitude due to the wall fluctuation modes is also 
influenced by the value of $\epsilon_*$ in the 
same manner as that due to the continuous modes.
In particular, the contribution of the 
wall fluctuation modes is negligible for $\epsilon_*\gg1$
as shown in Fig.~\ref{fig:newfig}.

To summarize, in the context of the string theory landscape,
the current observational data already constrain the
shape of the potential such that the slow-roll parameter
$\epsilon=(V'/V)^2/(2\kappa)$ right after tunneling should be
large, $\epsilon_*\gg1$ unless $1-\Omega_0$ is extremely small.
A large value of $\epsilon_*$ also reduces the contribution
of the wall fluctuation modes 
to the CMB spectrum, and hence seems to be even preferred from
the observed suppression of the CMB quadrupole.

In this paper, we focused on tensor-type perturbations.
We found that the effect of the evolution of the inflaton
inside a nucleated bubble affects the tensor-type power spectrum
significantly.
We also discussed expected signatures in the scalar-type perturbations
based on previous work in the literature. 
Using these expected features in the CMB spectra, 
we can say that we are already testing models of inflation
in the context of string theory landscape. 
To make more definite predictions and test the landscape,
detailed studies on the scalar-type perturbations are necessary.
We plan to come back to this issue in the near future.

\section*{Acknowledgments}

D.Y. would like to thank K.~Kamada, S.~Mukohyama, T.~Nakamura,
Y.~Nambu, Y.~Sendouda, L.~Susskind  and J.~Yokoyama 
for valuable comments and useful suggestions.
This work is supported in part by Monbukagakusho 
Grant-in-Aid for the global COE program, 
``The Next Generation of Physics, Spun from Universality 
and Emergence'' at Kyoto University.
The work of M.S. was supported by JSPS Grant-in-Aid for Scientific Research
(B) No.~17340075 and (A) No.~18204024 and by Grant-in-Aid for Creative Scientific 
Research No.~19GS0219.
T.T. is supported by Monbukagakusho Grant-in-Aid for
Scientific Research No. 21111006 and No. 22111507.
D.Y. and A.N. were supported by Grant-in-Aid for JSPS No.~20-1117 and No.~21-1899, and
A.L. was supported in part by NSF Grant No. PHY-0756174 and  by the FQXi Grant No. RFP2-08-19.


\end{document}